\documentclass[prl,twocolumn,preprintnumbers,amsmath,amssymb,showpacs]{revtex4}

\usepackage[usenames]{color}
\usepackage{graphicx}
\usepackage{amsfonts}
\usepackage{amssymb}
\usepackage{epsfig}
\usepackage{bm}
\usepackage{dcolumn}
\newcommand{\beq}{\begin{eqnarray}}
\newcommand{\eeq}{\end{eqnarray}}

\newcommand{\e}{{\text e}}

\newcommand{\bk}{{\bf k}}
\newcommand{\nk}{{\textrm{mom}}}
\newcommand{\ld}{{\textrm{dens}}}

\begin{document}
\title{Strongly Interacting Atom Lasers in Three Dimensional Optical Lattices}
\date{\today}

\author{Itay Hen}
\affiliation{Department of Physics, Georgetown University, Washington, DC 20057, USA}
\author{Marcos Rigol}
\affiliation{Department of Physics, Georgetown University, Washington, DC 20057, USA}

\begin{abstract}
We show that the dynamical melting of a Mott insulator in a three-dimensional lattice 
leads to condensation at nonzero momenta, a phenomenon that can be used to generate strongly 
interacting atom lasers in optical lattices. For infinite onsite repulsion, the case 
considered here, the momenta at which bosons condense is determined analytically and found to have 
a simple dependence on the hopping amplitudes. The occupation of the condensates is shown 
to scale linearly with the total number of atoms in the initial Mott insulator. 
Our results are obtained using a Gutzwiller-type mean-field approach, gauged 
against exact diagonalization solutions of small systems.
\end{abstract}

\maketitle

The invention of the first optical lasers more than half a century ago marked the beginning of 
the ultimate control over light waves in a way that would later revolutionize the world.
Now, with the realization of Bose-Einstein condensation (BEC) in atomic gases 
\cite{bec95}, we are entering an era in which a similar degree of control over 
matter waves is being achieved. In recent years, much effort has been devoted
to developing techniques for converting the trapped atoms of a BEC into freely propagating 
coherent matter waves -- the so called atom lasers \cite{s22,al2b}. The realization of 
atom lasers \cite{s22} has opened promising avenues of theoretical and 
experimental research. These coherent matter-wave beams are expected to become 
constituents in future scientific and technological devices \cite{al3}.

So far, most of the atom-laser experiments have dealt with weakly-interacting gases. 
The possibility of gaining further control by enhancing interactions by means of optical lattices has remained open. 
Here we address this issue and demonstrate that the free expansion of bosons from a Mott-insulating state 
into an empty lattice leads to the emergence of coherent matter waves at nonzero momenta, 
which in turn can be fully controlled by the hopping amplitudes in the lattice. In one dimension (1D), a 
related phenomenon was reported in Ref.\ \cite{s21}, where the melting of a Mott insulator
of hard-core bosons (HCBs) was shown to produce a quasi-coherent matter wave characterized by 
power-law decaying correlations. Interestingly, the power law was found to be the same as for 
the ground-state solution, i.e., in this transient regime, the system
cannot be characterized by a finite temperature. Analytical work in the $XY$ chain reproduced this 
behavior \cite{mitra10}, and further studies in 1D showed that soft-core bosons \cite{rodrig06}
and two-component fermionic systems \cite{fabian09} exhibit similar phenomena.
 
Following the promising results obtained for 1D systems, in this Letter we examine the dynamics of bosons
in higher dimensions. For concreteness, we will focus on the hard-core limit \cite{note0}.  
Studying the expansion in higher dimensions is of much interest not only because most 
experiments are performed in these regimes but also because true condensation, and hence a true 
atom laser, can only be realized in dimensions higher than one. We examine whether condensation
(off-diagonal long-range order) can develop dynamically during 
the expansion of an initial state that exhibits only short-range correlations, namely, 
a Mott-insulating state.

In this study, we consider HCBs on a three dimensional (3D) lattice, with $N=L_x \times L_{\perp}^2$ 
sites in the presence of a general onsite potential. Here, $L_x$ ($L_{\perp}$) denotes the 
number of sites in the $\hat{x}$ ($\hat{y}$ and $\hat{z}$) direction(s). The Hamiltonian can be written as:
\beq \label{hamil}
\hat{H}=-\sum_{ \langle ij \rangle } t_{ij} \left( \hat{a}_i^{\dagger} \hat{a}_{j} 
+ \hat{a}_{j}^{\dagger} \hat{a}_i \right) - \sum_i \mu_i \hat{n}_i \,,
\eeq
where $\langle ij \rangle$ denotes nearest neighbors, $\hat{a}_i$ ($\hat{a}_i^{\dagger}$) 
destroys (creates) a HCB on site $i$, $\hat{n}_i=\hat{a}_i^{\dagger} \hat{a}_i$ 
is the local density operator, $\mu_i$ is the local chemical potential, and the $t_{ij}$'s are the 
hopping amplitudes. The HCB creation and annihilation operators satisfy the constraints
$\hat{a}^{\dagger 2}_{i}= \hat{a}^2_{i}=0$, which preclude multiple site occupancies. Equation ~\eqref{hamil} 
may be viewed as a spin-1/2 $XY$ model with a site-dependent magnetic field applied in the $\hat{z}$ direction.

A typical initial state in our system is an almost perfect Mott insulator, created by placing HCBs 
in a very strong harmonic trap with curvature along the $\hat{x}$ direction, i.e., with $\mu_i=\mu_0-V x_i^2$, 
where $V$ is the curvature of the trap, $x_i$ is the distance of the $i$-th site from the center of 
the trap along the $\hat{x}$ direction, and $\mu_0$ is the chemical 
potential at the center of the trap. The resulting initial state is a `slab' of width $W_x$ around the center 
of the trap in the $\hat{x}$ direction with $\langle \hat{n}_i \rangle \approx 1$, 
which extends over the entire lattice in the transverse directions. 
In experiments, such a state could be produced by a box-like trap in the transverse directions and very 
strong on-site interactions ($U/t\gg1$). For simplicity, we consider periodic boundaries in the transverse
directions. This choice makes our system homogeneous in those directions as a result of which, the actual value of $L_{\perp}$ 
($L_{\perp}\geq2$) plays no role in the mean-field analysis.

At time $\tau=0$, the harmonic trap is turned off and the bosons start expanding freely throughout 
the lattice. Their particular initial state causes them to move in the $\hat{x}$ direction leaving 
the system translationally invariant in the transverse ones. We then study various 
properties of the system measured in the course of the evolution. 
Among those are density profiles, momentum distribution functions, 
and the condensate fraction along with the lowest natural orbital (LNO). 
The latter two are the largest eigenvalue of the one-particle density matrix 
$\rho_{ij}=\langle \hat{a}^{\dagger}_i \hat{a}_j \rangle$ and its eigenvector, respectively \cite{lno}.

Partial control over the dynamics of the system is gained by considering two different 
hopping parameters: The first is the hopping amplitude in the $\hat{x}$ direction, denoted by $t_x$. 
Without loss of generality, this amplitude is fixed at $t_x=1$, setting the energy scale. 
The second value corresponds to the hopping amplitude in the transverse directions $t_y=t_z$. 
For reasons that will become apparent later, we shall use the dimensionless quantity
$\eta=(t_y+t_z)/t_x$ as the control parameter.

In contrast to the special 1D case, where the model has an exact solution \cite{s21}, 
to explore the dynamics of HCBs in higher dimensions one must resort to approximate schemes.
This is because numerically-exact methods such as exact diagonalization, scale exponentially with 
the system size and hence allow insight into the dynamics of only small systems. 
Here, we probe the dynamics of HCBs primarily by means of a Gutzwiller-type mean-field 
approximation, adjusted to handle time evolution via the time-dependent variational principle 
method, originally developed for soft-core bosons \cite{mfa}. This approach becomes exact 
in the limit of infinite dimensions, but has proved to provide qualitatively 
correct phase diagrams for the ground state of two-dimensional (2D) and 3D 
hard-core boson systems \cite{hen0910}.

Within this method, we employ the following simple product ansatz as the wave-function of the system:
\beq \label{eq:ansatz}
| \psi \rangle =  \prod_{j=1}^{N} \e^{i \chi_j} \left(\sin \frac{\theta_j}{2} |0 \rangle + 
\cos \frac{\theta_j}{2} \e^{i \phi_j} | 1 \rangle   \right) \,,
\eeq
where we have utilized the usual parametrization of spin-1/2 particles.  
The polar angles $\theta_i \in [0,\pi]$, and the azimuthal angles, $\chi_i$ and 
$\phi_i\in [0,2 \pi)$, are time-dependent. By stationarizing 
$\langle \psi | i \partial /\partial \tau -\hat{H}| \psi \rangle$
(here, $\hbar=1$), we obtain the following set of first-order coupled equations:
\beq \label{eq:eomAnsatz}
\dot{\theta}_i = 2 \; \textrm{Im} \left[S_i\right] \,, \quad 
\dot{\phi}_i = \mu_i -2  \, \textrm{Re} \left[S_i\right] \cot \theta_i \,, 
\eeq
where ${S_i \equiv \frac1{2} \sum_j t_{ij} \sin \theta_j \e^{i (\phi_j-\phi_i)} \delta_{i,n(j)}}$, 
and  $\delta_{i,n(j)}$ is unity if $i$ and $j$ are neighbors
and zero otherwise \cite{note1}. 

Equations (\ref{eq:eomAnsatz}) are then solved simultaneously for all sites using a multi-dimensional 
forth-order Runge-Kutta method with an $\mathcal{O}(\Delta \tau^5)$ error, where $\Delta \tau$ 
is the discrete time step (taken to be $\Delta \tau=0.001$). Self consistency of the method is 
verified by monitoring the conservation of quantities such as total number of particles, 
total energy and total momentum. For the chosen time step, these quantities were observed to remain 
constant (up to 8 significant digits) throughout the entire simulation.  

Since our description of the model is approximate in nature, in order to gain an understanding 
of the accuracy of our results, we compare them against exact solutions. The latter are 
available in 1D, where the mean-field approach is not expected to be accurate \cite{supplement}, and for small 2D systems
where exact diagonalization calculations can be used. 

\begin{figure}[htp!]
\includegraphics[angle=0,scale=1,width=0.475\textwidth]{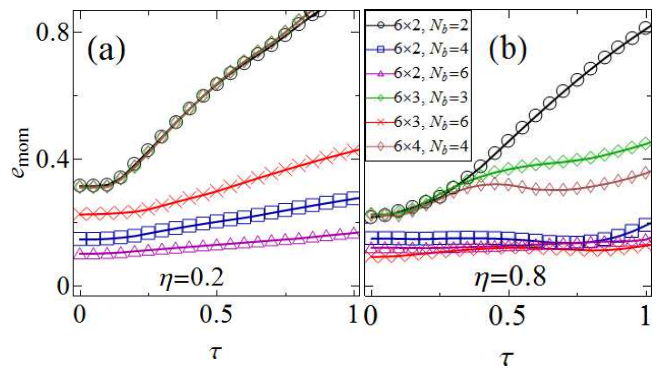}
\vspace{-0.3cm}
\caption{\label{fig:fig2derrs} (Color online) Momentum-distribution error $e_{\nk}$ [Eq.\eqref{eq:errNk}] as 
a function of time $\tau$ for different small 2D systems of size $L_x\times L_{\perp}$ with $N_b$ bosons. 
The hopping ratios are: (a) $\eta=0.2$ and (b) $\eta=0.8$ \cite{supplement}.}
\end{figure}

In Fig.\ \ref{fig:fig2derrs}, we present an analysis of several small 2D systems. 
The figure shows the time evolution of the normalized difference for $\langle \hat{n}_{\bk}\rangle$ between 
the mean-field results and the exact ones, defined as 
\beq \label{eq:errNk}
e_{\nk} =\frac{\sum_{\bk} \Big| \langle \hat{n}_{\bk}\rangle_{\textrm {\small exact}} - 
\langle \hat{n}_{\bk}\rangle_{\textrm {\small mean-field}} \Big|\,}
{\sum_{\bk} \langle \hat{n}_{\bk}\rangle_{\textrm {\small exact}}},
\eeq 
for lattices with up to $6\times4$ sites and up to six bosons, and two different values 
of $\eta$ \cite{supplement}. 

As expected, the errors increase during the expansion and attain significant values
in all cases ($\approx 0.8$ in the worst case and $\approx 0.15$ in the best case) even 
before they start bouncing off the edges of the lattice (which occurs at $\tau\sim1$). However, 
several conclusions may be drawn from comparison of the errors in the various cases. First, 
larger values of $\eta$ yield smaller errors [Fig.\ \ref{fig:fig2derrs}(a) vs 
\ref{fig:fig2derrs}(b)]. This is not surprising considering that as $\eta$ increases, the system 
moves from 1D ($\eta=0$) to `full' 2D ($\eta=1$), and that mean-field theories 
are expected to become more accurate as the dimensionality of the system increases \cite{hen0910}. 
Figure \ref{fig:fig2derrs} also shows that two other factors contribute to the reduction of the 
error, (i) adding particles to the system, and (ii) increasing the lattice size in the 
transverse direction \cite{supplement}.

The fact that in 2D, the mean-field results are qualitatively similar to the ones obtained with 
exact diagonalization \cite{supplement}, and the quantitative decrease of the relative errors 
as $\eta$ becomes large or as the number of particles increases, suggest that in 3D one can not 
only gain a qualitative understanding of the dynamics of the bosons during the melting of a Mott insulator, 
but also an actual quantitative picture of it. This is both because mean-field is expected to be more 
accurate in 3D than in 2D, and because the systems we can simulate contain far more particles on much 
larger lattices. 

\begin{figure}[htp!]
\includegraphics[angle=0,scale=1,width=0.475\textwidth]{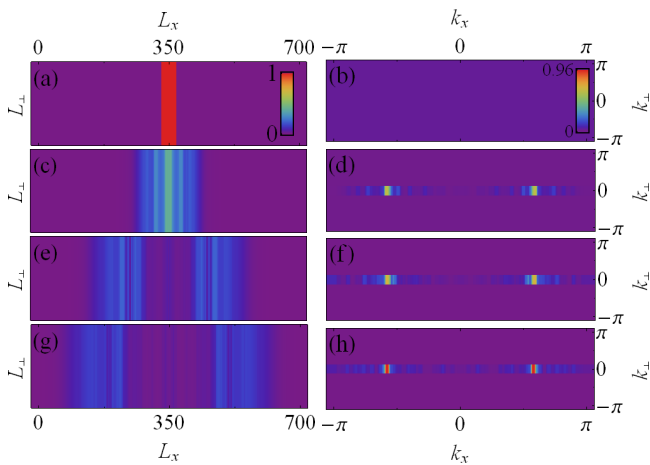}
\vspace{-0.3cm}
\caption{\label{fig:main} (Color online) Evolution of: density profiles (left panels) and momentum distribution function, 
as a fraction of $L^2_{\perp}$ with the $\hat{z}$ direction omitted (right panels), during the free expansion of 
$40\times L^2_{\perp}$ HCBs in a 3D lattice with $L_x=700$ and $\eta=0.8$. The times shown here are, from top to bottom,
$\tau=0$, $\tau=84$, $\tau=151$ and $\tau=187$.}
\end{figure}

Figure \ref{fig:main} depicts the main result of this Letter. It shows the density profiles and momentum 
distribution functions of the evolution of $N_b=40\times L^2_\perp$ bosons on a $700\times L^2_\perp$-site lattice, 
during the melting of a three-dimensional Mott insulator with $\eta=0.8$. Several important features are apparent 
in these plots. In the density profiles (left panels), one can see that during the expansion, after the 
insulator has completely melted, two solitonic `lumps' of bosons form and subsequently move in opposite 
directions with constant velocity. Even though the lumps slowly broaden as they travel, they retain their shape 
[compare Figs.~\ref{fig:main}(e) and \ref{fig:main}(g)]. Cross-sectional plots of Fig.\ \ref{fig:main} are presented 
in the supplementary material. 

The above behavior of the density profiles is accompanied by an even more remarkable
behavior of the momentum distribution function (right panels in Fig.\ \ref{fig:main}). While 
$\langle \hat{n}_{\bk}\rangle$ of the initial state is almost completely flat (as the Mott 
insulator is localized in real space), as the system expands, interactions between bosons redistribute
the energy. This results in the emergence of two condensates at opposite momenta. Given the dispersion 
relation in the lattice, these momenta determine the velocity of the 
two lumps seen in the density profiles \cite{supplement}.

As it turns out, the momenta at which condensation takes place can be evaluated analytically 
without resorting to mean-field theory. To calculate them, we first obtain the dispersion relation of the 
system in the absence of the trapping potential, given by  
\beq \label{eq:ek}
\epsilon_{\bk}\equiv\langle\bk|\hat{H}|\bk\rangle=-2 t_x \cos k_x-2 t_y \cos k_y-2 t_z \cos k_z\,
\eeq
where $|\bk \rangle = \hat{a}_{\bk}^{\dagger} |0 \rangle$ and $\hat{a}_{\bk}^{\dagger}$ 
($\hat{a}_{\bk}$) creates (destroys) a HCB with momentum ${\bf k}$. Note that even though 
the dispersion relation is the same as for noninteracting particles, HCBs do interact due to the 
constraints on site occupancy. (This is reflected in Fig.\ \ref{fig:main}, where momenta are 
redistributed during the expansion.) In our setup, the initial total energy of the system is 
$\epsilon\approx0$, as can be immediately verified by considering the energy of the insulating ground state. 
Since the expansion takes place in the $\hat{x}$ direction, it is expected that if condensation 
occurs it would take place at momenta $(\pm q_x,0,0)$, for some $q_x$ which is to be determined. 
Further assuming that the majority of the bosons occupy these two modes, together with conservation 
of energy, leads [from Eq.\ (\ref{eq:ek})] to a simple relation between the momentum of the 
condensate ($q_x$) and the parameters of the model:
\beq \label{eq:main1}
\cos q_x = -\eta  \,.
\eeq
This remarkable behavior of the HCB gas therefore implies that the locations 
of the two prominent momentum modes are fully controlled by the hopping parameters. 

\begin{figure}[!b]
\includegraphics[angle=0,scale=1,width=0.475\textwidth]{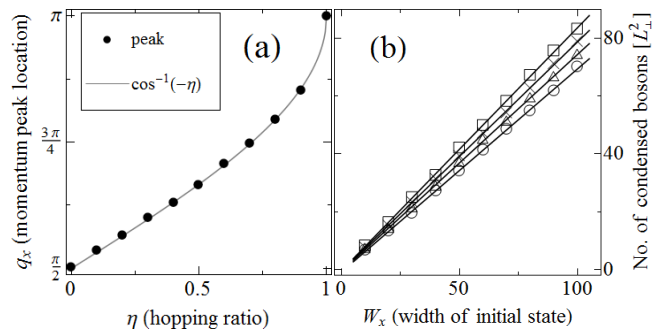}
\vspace{-0.3cm}
\caption{\label{fig:figScal} (a) Dependence of the momentum peak locations on the hopping parameters. 
The data points are taken from the various simulations whereas the line is
$\cos^{-1} (-\eta)$. (b) Scaling of the three-dimensional system: Total number of condensed bosons 
(in units of $L^2_{\perp}$) as a function of width of the initial Mott insulating state 
for $\eta=0.2$ ($\bigcirc$), $\eta=0.4$ ($\triangle$), $\eta=0.6$ ($\times$) and $\eta=0.8$ ($\square$).}
\end{figure}

The mean-field results agree in full with this analytical argument. This is shown in 
Fig.\ \ref{fig:figScal}(a) where we plot the locations of the momentum peaks for different values 
of $\eta$ as predicted by Eq.\ (\ref{eq:main1}) (solid line) and the result of 
the mean-field simulation (points). The velocity at which the condensates travel in the lattice 
is then given by ${\bf v}_{\textrm{g}} ={\bf \nabla} \epsilon_{\bf k} =2 t_x \sin q_x \hat{x}$. 
Interestingly, if the hopping parameters are such that the right-hand-side of Eq.\ (\ref{eq:main1}) 
is greater than unity, the bosonic gas does not melt within the mean-field theory. This is an artifact 
of the mean-field approach. Exact-diagonalization calculations on 2D systems show
that the Mott insulator does slowly melt for $\eta>1$. Hence, we expect that in large finite-dimensional 
systems with $\eta>1$ the Mott insulator will melt but no condensation will occur. 
Since the 2D systems we can study with exact diagonalization are too small to verify this 
hypothesis, this matter will have to be tested in experiments with ultracold bosons in optical 
lattices.

So far, we have been stating without proof that during the expansion of the initial Mott insulator, 
the emergence of peaks at nonzero momenta signals condensation. This was certainly not the case in the 
1D systems studied in Ref.\ \cite{s21}, even though similar peaks were observed at momenta $q_x=\pm \pi/2$. 
In 1D, the occupation of the two LNOs [which correspond to the two peaks in $n(k)$] was found to be proportional to $\sqrt{N_b}$, 
meaning that, in the thermodynamic limit, each condensate fraction (the ratio of the number of particles 
in each LNO to the total number of particles in the system) is zero \cite{s21}. True condensation only 
takes place if the LNO occupation is in proportion to the total number of particles in the system 
\cite{leggett01}, hence one must verify that it scales with $N_b$.

In Fig.\ \ref{fig:figScal}(b), we study the scaling of the total occupation of the LNOs 
in our systems for different values of $\eta$ ranging from $0.2$ to $0.8$. 
As the figure indicates, the number of condensed bosons indeed grows linearly with the width $W_x$, 
i.e., with the total number of bosons in the initial Mott insulating state. 
As the value of $\eta$ increases and one moves from a quasi-1D system to a more 3D system, 
the slope, that is, the fraction of bosons that condense, grows as well. While it is well known that true condensation can occur
in 2D systems at zero temperature and in 3D or higher dimensional systems below some critical 
temperature, the surprising result here is that condensation is found out of equilibrium, in a 
system where the energy is conserved, and for an insulating initial state that exhibits no off-diagonal correlations.

The dynamics of the systems analyzed above describes bosons hopping on a 3D lattice with periodic boundary 
conditions in the transverse directions. These were chosen in order to eliminate edge effects and to keep the system 
homogeneous in those directions. This geometry is, unfortunately, not realizable experimentally. In experiments, 
the boundaries are naturally open (a box trap or other trapping potentials). In the supplementary material,
we show that our conclusions are not altered when open boundaries (a box trap) are considered in the
transverse directions provided that $L_{\perp}$ is sufficiently large. This is to be expected since  
$L_{\perp}\rightarrow\infty$ and periodic boundary conditions in the $\perp$ directions are equivalent.
In practice, the results are found to be similar for $L_{\perp}\gtrsim40$ \cite{supplement}, which are lattice sizes
that are accessible experimentally.

In summary, we have shown that the melting of a Mott insulator with $\langle \hat{n}_i \rangle \approx 1$ in a three dimensional lattice
leads to the dynamical emergence of condensates at nonzero momenta. We have determined the momenta at which
condensation takes places and showed that these values have a simple dependence on the hopping amplitudes in the lattice.
We have also shown that the occupation of these out-of-equilibrium condensates scales with the 
number of particles in the initial Mott insulator, and with a proportionality constant that increases with 
$\eta$. We followed a mean-field approach that was gauged against exact diagonalization 
results of small 2D systems, where qualitatively similar results were obtained within both approaches.

Our results suggest that the expansion of a Mott insulator can be used to generate strongly interacting atom lasers 
in optical lattices, where the velocity of the condensates (and their wavelengths) can be fully controlled by the 
parameters of the optical lattice involved in the setup. This study thus presents evidence for a fundamentally new 
phenomenon, namely, out-of-equilibrium condensation in an isolated, and hence, energy-conserving system. As such, 
this study may have important implications on current experiments with ultracold Bose gas in optical lattices, 
as well as on our understanding of quantum systems out of equilibrium. We note here that experimental setups capable 
of verifying the results presented here are currently available \cite{weiss}. As mentioned before, we have checked that
this phenomenon also occurs in the soft-core boson case for large onsite repulsive interactions, in which case
the dispersion relation, and hence the momenta at which the bosons condense, becomes dependent on the value of 
the onsite repulsion. 

\begin{acknowledgments}
This work was supported by the US Office of Naval Research under Award No.\ N000140910966. We are grateful
to E. Khatami and F. A. Wolf for useful discussions.
\end{acknowledgments}

\pagebreak

\onecolumngrid

\vspace*{0.4cm}

\begin{center}

{\large \bf Supplementary material for EPAPS
\\ Strongly Interacting Atom Lasers in Three Dimensional Optical Lattices}\\

\vspace{0.6cm}

Itay Hen and Marcos Rigol\\

{\it Department of Physics, Georgetown University, Washington, DC 20057, USA}

\end{center}

\vspace{0.6cm}

\twocolumngrid

\section{Mean-field vs exact results in 1D}

\begin{figure}[!b]
\includegraphics[angle=0,scale=1,width=0.45\textwidth]{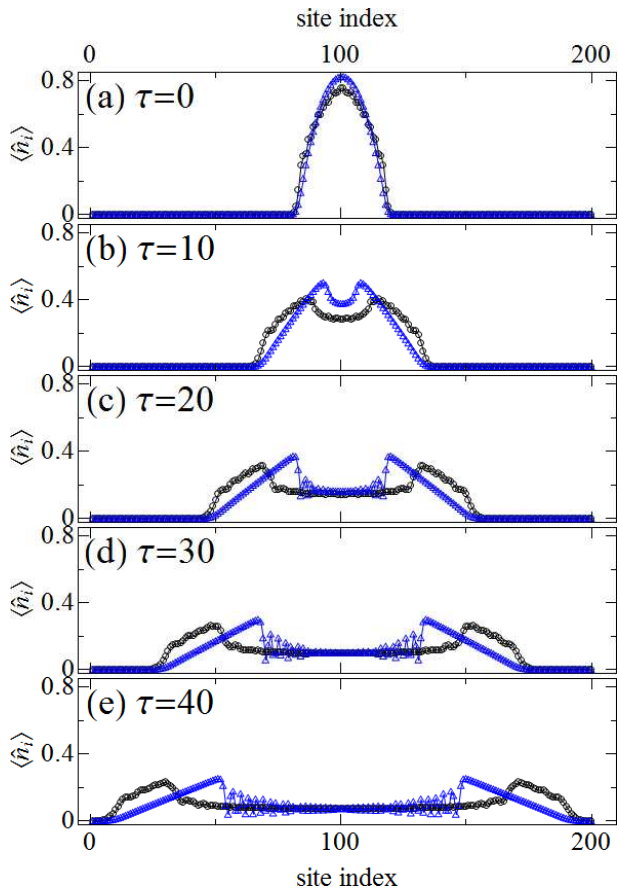}
\caption{\label{fig:fig1DMF} Evolution of 20 HCBs on a 200-site 1D lattice. 
The HCBs are initially trapped inside a harmonic potential and are then released. 
The figure shows the density profiles of the system at five different times (from top to bottom: $\tau=0, 10,20,30$ and $40$) 
as it was computed by exact means (circles) and by the mean-field approximation (triangles).}
\end{figure}

In one dimension, the existence of an exact solution to our model of interest (see Ref.\ [5] in the Letter) 
enables us to compare mean-field results against exact ones in systems with lattices which are comparable in size
with those realized experimentally. The 1D comparison should however be treated with caution 
because the mean-field approximation is expected to yield meaningful results only in higher 
dimensions. Nonetheless, in Fig.\ \ref{fig:fig1DMF} we show several snapshots of the density profile of a 1D system,
taken at five different times. These show the mean-field local densities (triangles), compared against matching exact ones (circles).
In both cases, 20 bosons are released from a trap with $V=0.01 t_x$ on a 200-site lattice.

It is clear from the figure that despite the visible quantitative differences between the mean-field 
evolution and the exact one, the two systems exhibit a qualitatively similar expansion. In both cases, two 
`lumps' of bosons emerge during the melting of the Mott insulator, and subsequently move away from each other 
at the same constant velocities. Remarkably, the fronts of the exact and mean-field condensates 
propagate together, as can be verified by examining the points where the density goes to zero in both solutions.

\begin{figure}[!h]
\includegraphics[angle=0,scale=1,width=0.35\textwidth]{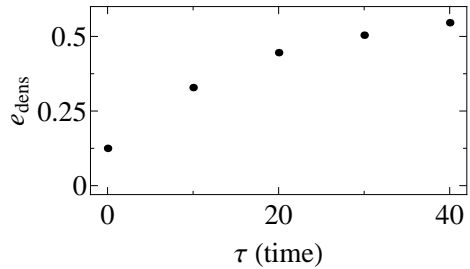}
\caption{\label{fig:errMF} The error $e_{\ld}$ [Eq.\ (\ref{eq:err})] as a function of time for the 
mean-field evolution displayed in Fig.\ \ref{fig:fig1DMF}.}
\end{figure}

Figure\ \ref{fig:errMF} shows the normalized difference between the mean-field local densities and the exact ones, 
given by: 
\beq \label{eq:err}
e_{\ld} =\frac{\sum_{i} \Big| \langle \hat{n}_i\rangle_{\textrm {\small exact}} - 
\langle \hat{n}_i\rangle_{\textrm {\small mean-field}} \Big|
}{\sum_{i} \langle \hat{n}_i\rangle_{\textrm {\small exact}}}\,.
\eeq 
The errors clearly increase with time. However, they do not increase as fast as may have naively been expected 
for a mean-field solution applied to a 1D system. This is encouraging in terms of the relevance of the 
mean-field approach for this model when applied to higher dimensional systems.

\section{Mean-field vs exact results in 2D}

Here, we discuss in further detail the exact setup through which the analysis of the momentum-error 
presented in the Letter [see Eq.\ (4) there] was performed. We also extend the analysis by considering 
the behavior of the error for two additional values of the parameter $\eta$ and discuss a similar analysis performed 
on the errors of the local densities [Eq.\ (\ref{eq:err})]. This is done in order to 
strengthen the conclusions drawn in the text in that context.

\begin{figure}[!h]
\includegraphics[angle=0,scale=1,width=0.475\textwidth]{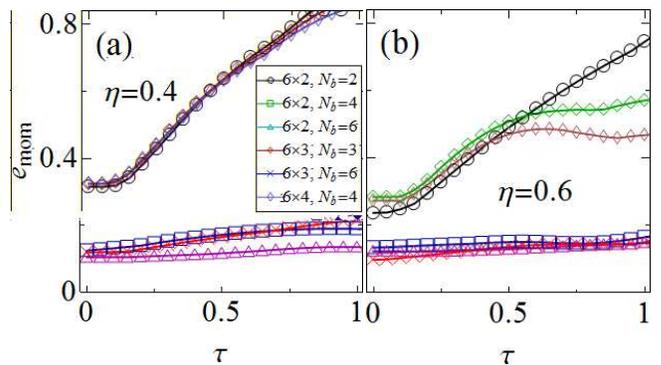}
\caption{\label{fig:fig2derrsB} (Color online) Momentum-error $e_{\nk}$ [Eq.~(4) in the Letter] as 
a function of time for different small 2D systems of size $L_x\times L_{\perp}$ with $N_b$ bosons. 
The hopping ratios are: (a) $\eta=0.4$ and (b) $\eta=0.6$. As the figures indicate,
the errors in the $\eta=0.6$ case are smaller, since larger values correspond to systems which 
are more two-dimensional. Also, as the number of particles in the system or the number of lattice sites 
in the transverse direction increase, the errors become smaller. These results are in accord with 
those shown in Fig.\ 1 in the Letter.}
\end{figure}

\begin{figure}[htp!]
\includegraphics[angle=0,scale=1,width=0.45\textwidth]{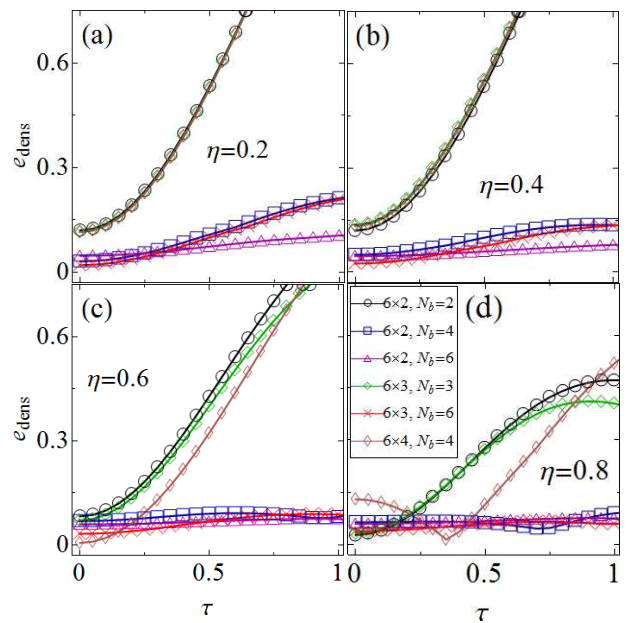}
\caption{\label{fig:fig2derrsC} (Color online) The error $e_{\ld}$ [Eq.\ (\ref{eq:err})] as a function 
of time $\tau$ for different small 2D systems of size 
$L_x\times L_{\perp}$ with $N_b$ bosons. The hopping ratios are: 
(a) $\eta=0.2$, (b) $\eta=0.4$, (c) $\eta=0.6$, and 
(d) $\eta=0.8$.}
\end{figure}

Since in 2D, exact diagonalization methods allow insight into the dynamics of systems with only a small number 
of particles hopping on small-size lattices, the setup for the error analysis presented here is slightly different 
than the basic scenario discussed in the Letter where bosons were released from a trap whose origin was the center 
of a periodic lattice. Here, we have used a harmonic potential along the $\hat{x}$-direction centered around one edge of the lattice, supplemented 
by open boundary conditions in the $\hat{x}$ direction. This setup takes advantage of the parity symmetry ($x \to -x$)
of the expansion observed in the conventional setup. This method has been used in the time-dependent DMRG calculations 
in Ref.~[8] in the Letter. In order to make the behavior of the different systems comparable,
the characteristic density (see Ref.~[5] in the Letter) of the different initial states was kept fixed for all 
the systems examined. This was done by choosing the curvature of the harmonic potential to be $V=V_0/W_x^2$ 
(in the figures, $V_0=2$, and $W_x=1,\,2$, or $3$). In Fig.~1 in the Letter, and in 
Figs.~\ref{fig:fig2derrsB} and \ref{fig:fig2derrsC} here, the behavior of the error is shown up to $\tau=1$, which is 
the typical time it takes for the particles to reach the opposite edge of the lattice and bounce back. 

In the Letter, we have shown that, in general, as the value of $\eta$ increases, 
the discrepancies between the mean-field and exact momentum profiles decrease. 
There, this was shown for two values of $\eta$, namely, $\eta=0.2$ and $\eta=0.8$.
To complete this picture and to provide further corroboration of our conclusions, 
in Fig.\ \ref{fig:fig2derrsB} we display the behavior of the momentum-error as a function 
of time for the same small 2D lattices but with $\eta=0.4$ and $\eta=0.6$.

In accord with the conclusions reported in the text, it can be seen in Fig.\ \ref{fig:fig2derrsB} 
that larger values of $\eta$ yield smaller errors [Fig.\ \ref{fig:fig2derrsB}(a) 
vs \ref{fig:fig2derrsB}(b)], but also that adding particles and/or sites in the transverse direction both reduce the errors as well. 
The two figures fit with the trend set by the  $\eta=0.2$ and $\eta=0.8$ panels of Fig.\ 1 in the Letter,
and therefore strengthen our previous argument that for many-particle 3D systems the mean-field
approach may be a good approximation for the study of the dynamics of 3D bosons.

Further support to our conclusions is provided by the behavior of the errors of the local densities. 
These are computed via Eq.\ (\ref{eq:err}), and depicted in Fig.\ \ref{fig:fig2derrsC} for the four different values
of $\eta$. Here too we have found that the errors decrease with increasing $\eta$, 
and that increasing the lattice size in the transverse direction and the addition of particles both reduce 
the difference between the mean-field and exact results. 

\section{Cross sections}
\begin{figure}[htp!]
\includegraphics[angle=0,scale=1,width=0.475\textwidth]{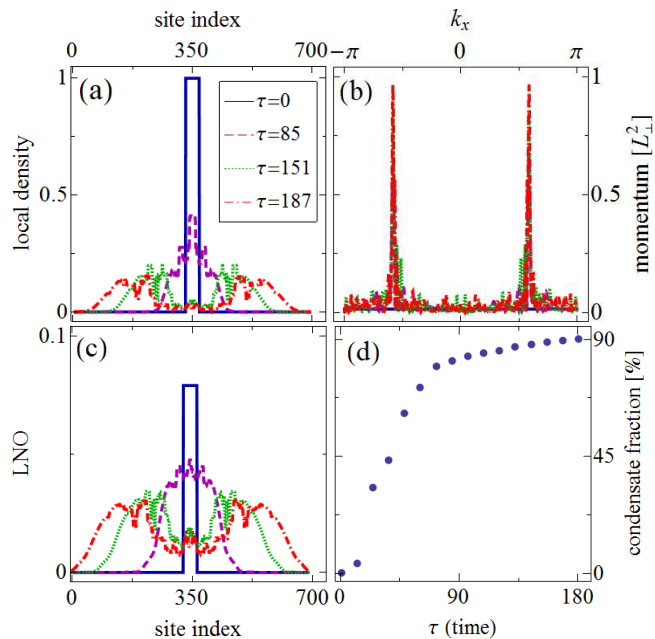}
\caption{\label{fig:figCross} (Color online) Evolution of (a) density profiles
(b) momentum profiles (c) absolute value of the LNO wave function during the free expansion of a Mott insulator
with $W_x=40$ in a 3D lattice with $L_x=700$ and $\eta=0.8$. The times shown here are $\tau=0$, $\tau=84$, $\tau=151$ and $\tau=187$.
(d) Total number of condensed bosons (LNO occupation) in the system divided by the total number of bosons,
as a function of time.  
}
\end{figure}

In Fig.\ 2 of the Letter, we presented intensity plots of the density and momentum profiles
of the bosonic gas during the expansion of a Mott insulator in 3D. In Fig.\ \ref{fig:figCross}, we provide 
the cross sections of the same density and momentum profiles, with cuts along the $x$ and $k_x$ axes,
i.e., $y=z=0$ and $k_y=k_z=0$, respectively. 

As evident from Fig.\ \ref{fig:figCross}(a), when the initially-trapped bosonic gas is released, 
two solitonic lumps begin to form during the melting of the Mott insulator. Once the latter has disappeared,
the lumps start moving away without a considerable change in their shape while they slowly broaden. 
The cuts across the momentum distribution function [Fig.\ \ref{fig:figCross}(b)] show very sharp peaks
emerging at finite momenta. Those peaks signal condensation, as discussed in the text, where the lowest 
natural orbital occupation was found to scale proportionally with the number of particles in the system.

In Fig.~\ref{fig:figCross}(c), we show the cross section along the $x$ axis of the lowest natural orbital.
The figure illustrates that the condensates are mainly localized in, and have similar shapes as, the two lobes visible 
in the density profiles. This explains why the (constant) group velocity of the condensates, determined by 
their momenta and the dispersion relation in the lattice (see related discussion in the Letter), is equal to the velocity of 
motion of the lobes appearing in the density profiles. 

The condensate fraction, i.e., the ratio between the occupation of the LNO  
and the total number of bosons, is shown in Fig. 5(d) as a function of time. 
There, one can see that it quickly increases in the initial stages of the 
expansion, during the melting of the Mott insulator, and then, after the 
insulator has melted, it exhibits a slow increase during the remainder of the 
evolution. The number of condensed bosons at the kink connecting the two regimes 
is the one reported in Fig. 3(b) of the Letter. Interestingly, while the 
expansion of the Mott insulator in our selected geometry can only cause the 
emergence of two independent condensates traveling in opposite directions 
(during their formation, the insulating slab between them precludes any phase 
correlations), the diagonalization of the one-particle density matrix computed 
within the mean-field approximation yields only one large eigenvalue, i.e., 
mean-field merges the two LNOs into one. Hence, the mean-field LNO occupation 
reflects the total number of condensed bosons (that is, the sum of the 
occupations of the two moving condensates). This can be easily verified
by examining the evolution of only one half of the system described in the 
previous section.
  
\section{Effects of open boundary conditions}

\begin{figure}[hbp!]
\includegraphics[angle=0,scale=1,width=0.45\textwidth]{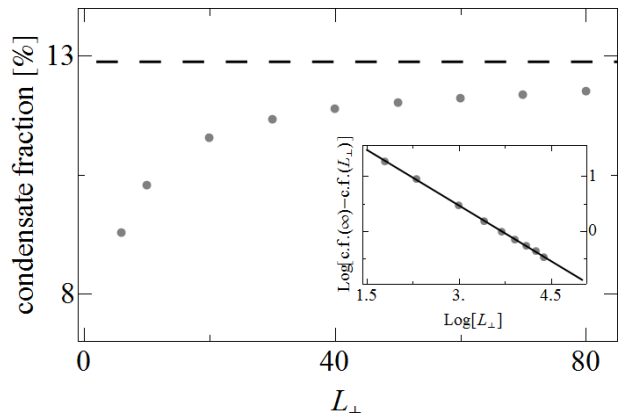}
\caption{\label{fig:figOpen} Effects of open boundary conditions. Condensate fraction (given in $\%$) at $\tau=18$
as a function of lattice size in the transverse direction, for a 2D system with $L_x=100$ and $\eta=0.2$. 
As the size of the system in the transverse
directions increases, the condensate fraction approaches the infinite-size (periodic) value marked by 
the dashed horizontal line. The inset shows a linear fit on a log-log scale illustrating 
the power-law behavior of the condensate fraction as it reaches the periodic limit. 
The exponent here is $\approx -0.675$.}
\end{figure}

In order to verify that open boundary conditions in the transverse direction do not alter 
the general features of the systems studied in the Letter, we have tested the effects of 
open boundaries on the condensate occupation in the lattice as a function of the transverse linear 
system size $L_{\perp}$. This was done in order to illustrate that in large systems edge effects 
do not destroy the general phenomenon described earlier. The results of this investigation are summarized 
in Fig.\  \ref{fig:figOpen}. The data points represent the condensate fraction at a fixed time 
($\tau=18$) after the trap has already been removed and the bosons were allowed to expand and condense. 

We have tested the effects of the open boundaries on the condensation in several 2D systems all with $100$ sites 
in the $\hat{x}$ direction but with varying $L_{\perp}$ (here, $\eta=0.2$). The dashed horizontal line marks the 
same quantity in the periodic ($L_{\perp} \to \infty$ limit) case discussed in the Letter. 
As expected, the larger $L_{\perp}$ is, the less pronounced edge effects become and we can therefore 
conclude that provided $L_{\perp}$ is large enough, systems with open boundary
conditions (a box trap) in the transverse direction will exhibit the  exact same features of
parallel periodic systems. This will remain true in the time scales relevant for the expansion of the bosons 
or the formation of the solitonic condensates.
The inset in the figure is a log-log scale of the data points, illustrating that the condensate
occupation approaches the periodic value algebraically, where the exponent here is $\approx -0.675$.

\begin{thebibliography}{100}
\bibitem{note0}
Qualitatively similar results were found for strongly interacting soft-core 
bosons and will be presented elsewhere.
\bibitem{note1}
The global phases $\chi_i$ play no physical role here, as they do not enter into the `physical' 
$\theta_i$ and $\phi_i$ equations.
\end{thebibliography}
\end{document}